\documentstyle[pre,aps,epsf,twocolumn]{revtex}
\begin{document}

\title{Damage spreading for one-dimensional, non-equilibrium models with parity 
conserving phase transitions}
\author{G\'eza \'Odor}
\address{\it Research Institute for Materials Science, 
H-1525 Budapest, P.O.Box 49, Hungary }
\author{N\'ora Menyh\'ard}
\address{\it Research Institute for Solid State Physics,
H-1525 Budapest, P.O.Box 49, Hungary}

\address{
\centering{
\medskip\em
\begin{minipage}{14cm}
{}~~The damage spreading (DS) transitions of two one-dimensional stochastic 
cellular automata suggested by Grassberger (A and B) and the kinetic Ising 
model of Menyh\'ard (NEKIM) have been investigated. 
These non-equilibrium models exhibit non-directed percolation universality 
class continuous phase transition to absorbing states, exhibit parity 
conservation (PC) law of kinks and have chaotic to non-chaotic DS phase 
transitions, too. 
The relation of the critical point and the damage spreading point has been 
explored with numerical simulations. For model B the two transition points
are well separated and directed percolation universality was found both for 
spin damage and kink damages in spite of the conservation of damage 
variables modulo 2 in the latter case. For model A and NEKIM  the two 
transition points coincide with drastic effects on the damage of spin 
and kink variables showing different time dependent behaviours. 
While the kink DS transition of these two models shows 
regular PC class universality, the spin damage of them exhibits a 
discontinuous phase transition with compact clusters and PC-like spreading 
exponents. In the latter case the static exponents determined
by finite size scaling are consistent with that of the spins of the 
NEKIM model at the PC transition point. The generalised hyper-scaling law 
is satisfied.
Detailed discussion is given concerning the dependence of DS on initial 
conditions especially for the A model case, where extremely long relaxation 
time was found.
\end{minipage}
}}
\maketitle

\section{Introduction}

While damage spreading (DS) was first introduced in biology \cite{Kaufman} it has 
become an interesting topic in physics as well \cite{Creutz,Stanley,Derrida}.
The main question is if a damage introduced in a dynamical system survives or 
disappears. To investigate this the usual technique is to make replica(s) of the
original system and let them evolving with the same dynamics and external noise.
This method has been found to be very useful to measure accurately dynamical
exponents of equilibrium systems \cite{GrasA}. It has turned out however, that the
DS properties do depend on the applied dynamics. An example is the case of the
two-dimensional Ising model with heath-bath algorithm versus Glauber dynamics
\cite{Mariz,Jan,GrasJPA}.

To avoid the dependences on dynamics Hinrichsen et al.\cite{HinWD} 
suggested a definition of 
"physical" family of DS dynamics according to which the active phase may be divided
to a sub-phase within which DS occurs for every member of the family, another sub-phase 
where the damage heals for every member of the family, and a third possible sub-phase
where DS is possible for some members and the damage disappears for other members.
The family of possible DS dynamics is defined such that it is to be consistent with
the physics of the single replicas (symmetries, interaction ranges etc.).

The universality of continuous DS transitions is an other open question. There is a 
hypothesis raised by Grassberger \cite{GrasJ}, that damage spreading transitions 
generically belong to
the universality class of directed percolation (DP) if they are separated from the
ordinary critical point. This claim is based on the DP hypothesis applied for 
the absorbing type DS transition since we can consider the difference of the 
replicas as an other dynamical system evolving by a complex rule. 
According to the DP hypothesis -- conjectured first in the early eighties 
\cite{Jansen,GrasZ} -- in the absence of conservation laws every continuous 
phase transition of a system with scalar order parameter and local interactions to a
single absorbing phase would belong the universality class of the DP. 
There are other more complex models like those with several absorbing states \cite{PCP}
and multi-component systems \cite{multicomp}, which exhibit DP transition 
too. During the years the DP universality class has been proven to be extremely robust. 
For a long time only a few number of exceptions has been found, which don't 
belong to the DP class. These are the parity conservation (PC) models and the 
multiplicative noise systems \cite{MultiN}.

The first examples of the PC models were Grassberger's (A and B) stochastic cellular
automata (SCA) \cite{GrasAB}. The kinks ('$00$' and '$11$'-s) of these models
exhibit mod 2 parity conservation and the absorbing state is doubly degenerated. 
Following that a series of models in the same 
universality class have been discovered. In case of the branching annihilating random 
walk with even number of offsprings (BAWe) \cite{Taka,PCD} the parity of the 
"particles" is conserved and there is a single absorbing state. 
In the non-equilibrium Ising model with combined spin-flip and spin exchange dynamics 
(NEKIM) \cite{Nora,DRS} the kinks have local parity conserving symmetry as well as the 
absorbing state is symmetrically doubly degenerated. The three species monomer-monomer 
\cite{Bassler} (3MM) and the interacting monomer-dimer (IMD) models \cite{Park,ParkDP} 
are multi component models with parity conservations and symmetric absorbing phases.

The common feature of these models that force them in a non-DP universality
class was first conjectured to be the parity conservation (the PC class name
comes from here). 
However this had to be refined, because models were found with global parity
conservation but DP class phase transition (see example ref.\cite{Inui} (ISCA)).
Field theoretical investigations of the BAW models showed that the BAWe parity
conservation dynamics in one-dimension results in a new non-DP fixed point
possessing the PC class universality, while for for odd number of offsprings
(BAWo) the transition is in the DP class \cite{UweCardy,CardyUwe}.
Furthermore, among the multiple absorbing state models one can have DP behaviour 
and BAWe parity conservation dynamics together if the symmetry of the absorbing 
states is broken (see \cite{ParkDP} in case of IMD, \cite{meod} in case of the
NEKIM model and \cite{HinPC} in case of the Generalized Domany-Kinzel SCA (GDK)). 
This implies that for multi-absorbing state models the BAWe parity conservation 
is not a sufficient condition to have non-DP universality class but the symmetry 
of the ground state is necessary too. 

\bigskip
\centerline{Summary of PC related models.}
\centerline{The notation "k" refers to kink, "s" refers to spin,}
\centerline{"h" means external field as variable of the model.}
\bigskip
\centerline{
\begin{tabular}{|c|c|c|c|c|}\hline
Model   &Abs. state & Dynamics  &Univ.       &Ref. no.\\ \hline\hline
A-k	&$Z_2$ symm.& BAWe	& PC         &\cite{GrasAB} \\ \hline
B-k	&$Z_2$ symm.& BAWe	& PC         &\cite{GrasAB} \\ \hline
BAWe-s  & singlet  & BAWe	& PC	     &\cite{Taka,PCD,UweCardy,CardyUwe} \\ \hline
BAWo-s  & singlet  & BAWo	& DP	     &\cite{Taka,PCD,UweCardy,CardyUwe} \\ \hline
NEKIM-k &$Z_2$ symm.& BAWe	& PC         &\cite{Nora,meod,meor} \\ \hline
3MM-s   &$Z_2$ symm.& BAWe	& PC         &\cite{Bassler} \\ \hline
IMD-s   &$Z_2$ symm.& BAWe	& PC         &\cite{Park,ParkDP} \\ \hline
GDK-s   &$Z_2$ symm.& BAWe	& PC         &\cite{HinPC} \\ \hline
ISCA-s  &singlet    & Global-PC	& DP	     &\cite{Inui} \\ \hline
IMD-s+h &$Z_2$ broken& BAWe	& DP	     &\cite{ParkDP} \\ \hline
NEKIM-k+h&$Z_2$ broken& BAWe	& DP	     &\cite{meod} \\ \hline
GDK-s+h &$Z_2$ broken& BAWe	& DP         &\cite{HinPC} \\ \hline
\end{tabular}
}
\bigskip

A very recent study has shown that DS transition is possible in a 
one-dimensional non-equilibrium kinetic Ising model \cite{HD} too, 
and the universality class of the transition is not always in the DP class. 
The dynamics was engineered as the combination of two subrules such that  
it creates $Z_2$ symmetric passive states, the kink damage variables follow BAWe 
parity preserving dynamics and a PC universality class DS transition emerges.

In this work we have investigated the damage spreading behaviour of some 
one-dimensional PC models:
Grassberger's A and B stochastic cellular automata (SCA) \cite{GrasA} 
with synchronous dynamics and the NEKIM model \cite{Nora}. 
The 'kink' variables of these models possess parity 
conservation and continuous PC class phase transition.

In case of the NEKIM model the '$01$' and '$10$' pairs act as kinks and 
follow the basic (BAWe) elementary reactions :
\begin{itemize}
\item left-right diffusion,
\item{$X \to 3 X$ \ \ \ \ \  reproduction,}
\item{$2 X \to 0$ \ \ \ \ \  annihilation.}
\end{itemize}

Since at damage spreading problems we follow the evolution of two or more
replicas, we can consider it as a special, multi-component dynamical problem 
(here with multiple absorbing states too). 
Furthermore, when the DS point is inside the active phase 
there is a passive state of the damage variables with fluctuating replicas in the 
background. Therefore the simple DP universality hypothesis can not be applied here
(although this does not exclude a DP class phase transition). 

We also studied here the DS properties of spins of these systems, which can be 
regarded as the "dual" variables of kinks but they don't obey the parity conservation. 
The combined observation of spin and kink damage variables sheds light on the 
interplay of parity conservation, absorbing state symmetry and universality.
In this paper we shall show by numerical simulations that the universality here 
is determined by not only the dynamics but the symmetry of the absorbing 
state is a necessary condition again like in cases of \cite{ParkDP,meod}. 

If the critical point and the DS point coincide interesting things happen. 
While the kink damage exponents will belong to the PC universality class,
in case of the spin damage the static exponents determined by finite size scaling
are in agreement with that of the the pure NEKIM model at the PC transition 
point on the spin level \cite{meod}. Detailed discussion on this is forthcoming
\cite{cpc}.
 
The dependence of the DS results on the initial states of the replicas is discussed
because for model-A a very slow relaxation makes it an important point.

\section{The damage spreading simulation methods} 

The time dependent simulation is a well established method to locate critical points
and to measure dynamical critical exponents at the same time \cite{GrasTor}. 
Here we applied it for kink and spin damage variables for system sizes $L=4096-16384$
with periodic boundary conditions. A single spin-flip difference is introduced between 
two identical replicas at the beginning of each simulation runs. The difference of 
spin and kink variables is measured during a time evolution with identical rules 
and random numbers for both replica. The maximum number of simulation steps was 
chosen to be $t_{MAX}=L/2$, and so the damage variables can not reach the boundaries
and one can avoid finite size effects of them.
However simulating near the critical point causes long transients
hindering to see the true scaling behaviour within reachable times.

The role of initial states of the replicas is not discussed in DS simulation literature.
If the DS transition point is not in the neighbourhood of a critical point an 
exponentially quick transient to the steady state is expected, but if they coincide 
-- as in case of the Grassberger A model -- the evolution to steady state
slows down to power law time dependence and we can expect finite time effects. 
First random states have been chosen with equal and uniform 
distribution of $0$-s and $1$-s.
In case of model A SCA this resulted in very confusing results. Then we investigated 
the effects from starting with steady state configurations i.e. replicas were driven 
to steady state before the DS measurements.

The quantities characterising damage evolution 
show power like behaviour in the $t\to\infty$ limit at the 
damage spreading point ($p_d$) separating chaotic and non-chaotic phases. 
The Hamming distance will be the order parameter of this paper:
\begin{equation}
D(t) = \left < {1\over L} \sum_{i=1}^L \vert s(i) - s^,(i) \vert \right > \label{Dscal}
\end{equation}
where $s(i)$ may denote now spin or kink variables. Kink variables for these models are
the '$00$' and '$11$' pairs in case of the A,B SCA and '$01$' and '$10$' pairs for the
NEKIM model. If there is a phase transition point, the Hamming distance behaves 
in a power law manner at that point:
\begin{equation}
D(t)\propto t^{\eta} \ ,
\end{equation}
Similarly the survival probability of damage variables behaves as:
\begin{equation}
P_s(t)\propto t^{-\delta} \,
\end{equation}
and the average mean square distance of damage spreading from the center scales as:
\begin{equation}
R^2(t)\propto t^z \ .
\end{equation}

The evolution runs were averaged over $N_s$ independent runs for each different 
value of $p$ in the vicinity of $p_d$ ( but for $R^2(t)$ only over the surviving runs ).

To estimate the critical exponents and the transition points together
we determined the local slopes of the scaling variables. 
For example in case of the survival probability:
\begin{equation}
-\delta_p(t) = {\ln \left[ P_s(t) / P_s(t/m) \right] \over \ln(m)}
\end{equation}
and we have used $m=4$. 
In the case of power-law behaviour we should see a horizontal straight line as 
$1/t \to 0$, when $p = p_d$. The off-critical curves should possess curvature. 
Curves corresponding to $p > p_d$ should veer upward, curves with $p < p_d$ 
should veer downward.

The damage spreading measurement of the order parameter time scaling can be very 
effectively parallelised in a multi-spin code manner \cite{GrasJ}, since one 
needs only one random number for each site of the different replicas and so 
one can follow the evolution of $N_r = {32\times 31\over 2}$ replicas in a simple 
$32$-bit computer vector of length $L$. However this method is not applicable to 
measure the survival probability scaling and $z$, since the healing of differences 
among all the $N_r$ replicas takes a very long time and one can not introduce a 
single initial damage for each pair at the center of the lattice.
For the simulation of survival probability a very effective code has been 
implemented for a special, associative string processor 
\cite{ASP}.

To determine static exponents finite-size scaling (FSS) simulations were
performed as well. 
As it was shown by Aukrust et. al. \cite{Auk} and \cite{Jensen93}, 
FSS is applicable to continuous, non-equilibrium phase transitions.
At the critical point the order parameter steady state density ($D$)
and the fluctuation $\chi = L^d (<D^2> - <D>^2)$ scale with the system size as :
\begin{equation}
D(L) \propto L^{-\beta/\nu_{\perp}},  \label{cs}
\end{equation}
\begin{equation}
\chi (L) \propto L^{\gamma/\nu_{\perp}}, \label{fs}
\end{equation}
where $\nu_{\perp}$ is the correlation length exponent in the space direction :
\begin{equation}
\xi (p) \propto \vert p - p_c \vert^{-\nu_{\perp}} \ \ ,
\end{equation}
$\beta$ is the order parameter exponent in the steady state:
\begin{equation}
D(p) \propto \vert p - p_c \vert^{\beta} \ \ ,
\end{equation}
and $\gamma$ describes the fluctuation of it:
\begin{equation}
\chi (p) \propto \vert p - p_c \vert^{-\gamma}.
\end{equation}
Simulations were done in one dimension for lattice sizes : 
$L = 64, 128, ... ,2048$. 
The necessary time steps to reach steady state were determined experimentally.
The time evolution of the concentration was plotted, and the necessary time
steps was fixed for a given $L$ such as $200-500$ time values following the 
level off. Averaging was done for these $200-500$ values times the number of 
surviving samples ($500$). 
The $p_d$ values were taken from the time-dependent MC calculations.

The dynamic exponent $Z = \nu_{\|} / \nu_{\perp}$ can be determined from the FSS
of the characteristic time $\tau(p,L)$. In this study we measured the time necessary to
reach half of the steady state concentration starting from a single damage state. 
The characteristic time obeys the finite size scaling law :
\begin{equation}
\tau(p,L) \propto L^Z h((p-p_c) L^{1/\nu_{\perp}}) \ ,
\end{equation}
where $Z= \nu_{\|} / \nu_{\perp}$.
For this measurement we used the same damage concentration time evolutions  
as in case of the static runs above.

\section{The Grassberger B model}

A (BAWe) parity conserving dynamics can be realized on the kinks (or 'particles') 
of the following SCA (we show the configurations at $t-1$ and the probability of 
getting '$1$' at time $t$):
\begin{verbatim}
      t-1: 100 001 101 110 011 111 000 010
       t:   1   1   1   p   p   0   0   0
\end{verbatim}
The '$00$' and '$11$' pairs are the simplest kinks of the model.
The time evolution pattern for $p=0$ is a regular chess-board in $1+1$
dimension (Rule-50 with double degeneration), i.e. , the absorbing 
states are period-two antiferromagnetic.
For $p < p_c (=0.539(1))$ kinks disappear exponentially,
while for $p > p_c$ they survive with a finite concentration. 
In the $p=1$ limit we get the deterministic Rule-122, which is known to be chaotic.
So there is damage spreading phase transition besides the absorbing phase transition
of PC universality.

\subsection{Kink damage results}

First the simulations were started from two replicas of lattices with identical 
random initial states but with a 2-kink initial difference. 
The parity of the lattice forces even or (odd) number of 
initial kinks, therefore it is not possible to create odd numbered kink differences. 
The parity of kinks is conserved. The parity of kink differences (even) is conserved 
too. 
\begin{figure}[h]
  \centerline{\epsfxsize=8cm
                   \epsfbox{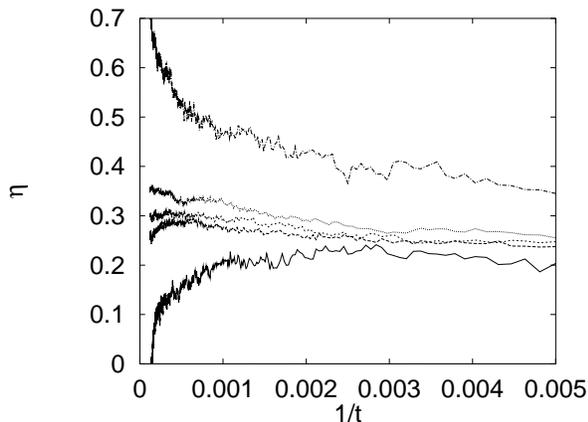}
  }
  \caption{Local slopes of the Hamming distance ($\eta$) in the B model, 
for $p=0.62,0.63,0.632,0.634, 0.65$ (curves from bottom to top). 
Statistical averaging was done over 10000 samples.}
 \label{fig11}
\end{figure}
Still we see a DP like universality of the damage variables 
(Fig. \ref{fig11},\ref{fig12}).
The location of the damage spreading point ($p_d = 0.632(1)$) is far from the 
PC critical point ($p_c = 0.539(1)$), therefore the active phase is divided to a
chaotic and non-chaotic sub-phase similarly to the case of the Domany-Kinzel 
SCA \cite{DKSCA,Martins,GrasJ}. 
\begin{figure}[h]
  \centerline{\epsfxsize=8cm
                   \epsfbox{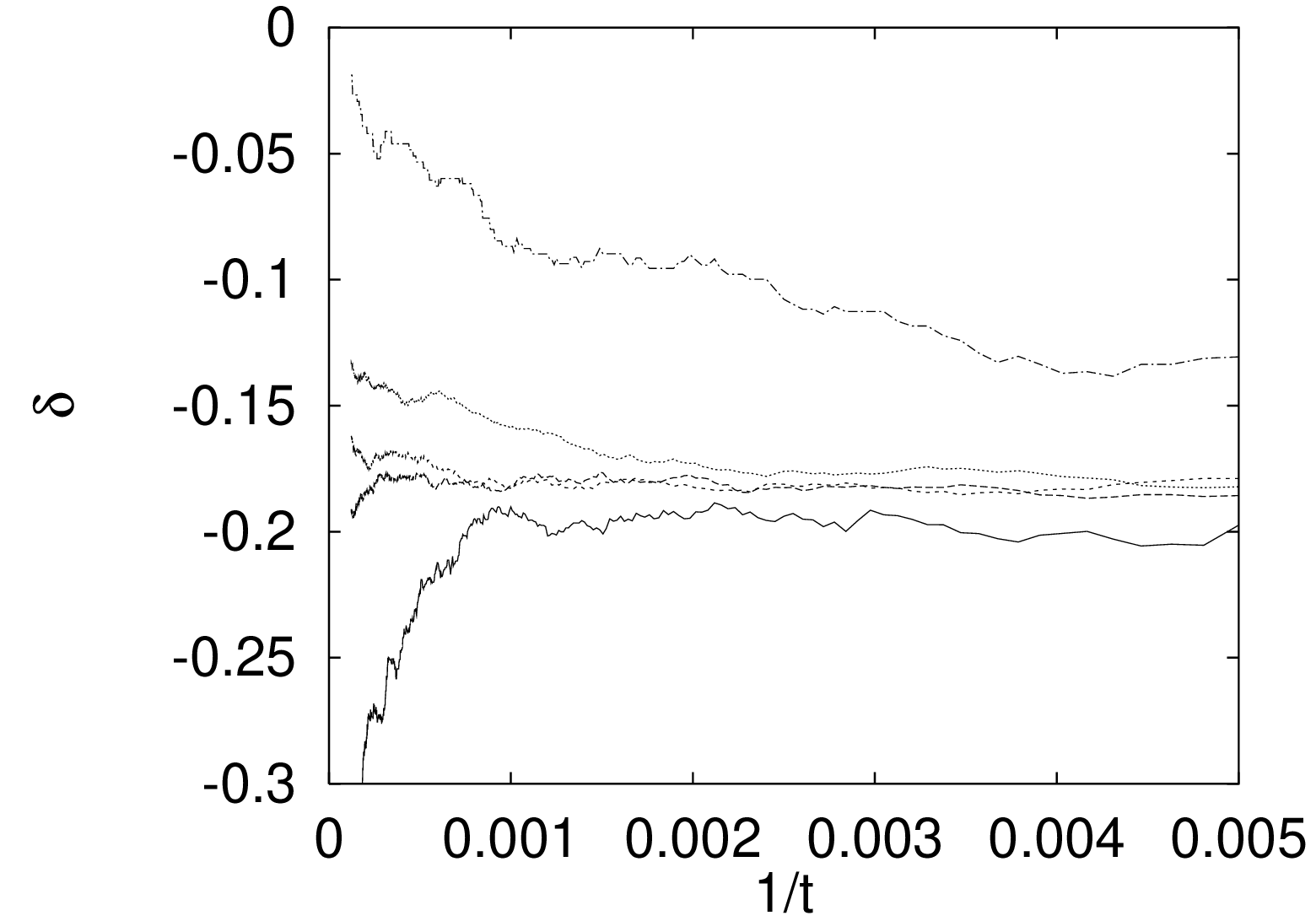}
  }
  \caption{Local slopes of the damage survival probability ($\delta$) in the B model, 
for $p=0.62,0.63,0.632,0.634,0.65$ (curves from bottom to top). 
Statistical averaging was done over 10000 samples.}
\label{fig12}
\end{figure}
The replicas at the DS point are in fluctuating states therefore they have not 
the "chess-board" double degeneration as in case of the PC critical point. 
The emergence of the DP exponents in spite of the mod 2 conservation of kink 
damage variables is similar to what was found numerically and analytically for 
PC models, when the external $H$ field destroyed the symmetry of the absorbing 
state \cite{ParkDP,meod,HinPC}. 
This suggests that the BAWe parity conservation rule is not a sufficient 
condition for having non-DP universality.

The same results has been found if we started the replicas from steady states 
with a single spin-flip initial difference.

\subsection{Spin damage}

Following the damage (difference) of the spins, instead of the kinks, we 
obtained the same DP like results. The universality was insensitive to the 
parity of the damage variables, it is DP for both cases.

\subsection{Finite size scaling for both cases}

\begin{figure}[h]
  \centerline{\epsfxsize=8cm
                   \epsfbox{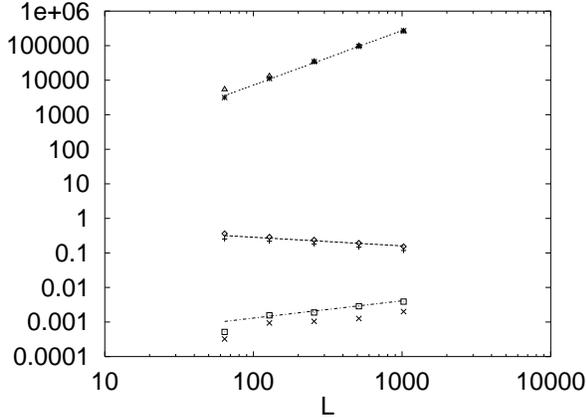}   
  }
  \caption{Finite size scaling results for the spin and kink damage for the B model.
   The diamonds correspond to the kink damage concentration, the squares to the
fluctuations of it. The crosses correspond to the spin damage concentration, the
"x"-s to the fluctuations of it. Triangles and stars denote characteristic times
$\tau$ of the spin and the kink damage cases.
Averaging was done over $500$ surviving samples.}
\label{grabfss} 
\end{figure}

Finite size scaling simulations were performed at the DS transition point
($p_d = 0.633$) for system sizes $L=64,128,...1024$. The necessary time steps 
to reach steady state were $t = 40000, 80000, ... $ respectively. 
The results can be seen on Figure \ref{grabfss}. 
As on can observe both the kink and spin damage concentration show a scaling with 
$-\beta/\nu_{\perp}=-0.25$, while the fluctuations have a slope 
$\gamma/\nu_{\perp}=0.5$, all agreeing with the corresponding exponents in the
DP universality class.
For the critical dynamical exponent $Z = \nu_{\|} / \nu_{\perp}$
DP-like scaling has been found for both the spin and kink damage cases.
Fitting can be done with $L^{1.5798}$ (DP value) in both cases.

\section{The Grassberger A model}

Another very similar model exhibiting parity conservation of kinks is the
Grassberger A stochastic cellular automaton:
\begin{verbatim}
      t-1: 100 001 101 110 011 111 000 010
       t:   1   1   0  1-p 1-p  0   0   1
\end{verbatim}
The time evolution pattern in $1+1$ dimension, for small $p$ 
evolves towards a stripe-like ordered
steady state (with double degeneration), while for $p > p_c (0.1245(5))$ the kinks 
(the '$00$' and '$11$' pairs survive. 
For $p=0$ we have the Rule-94, class 1 CA, while the $p=1$ limit is the chaotic
Rule-22 deterministic CA. Therefore we can expect a damage spreading phase transition
between $p = p_c$ and $p=1$ (damages always heal or survive in the ordered steady 
states). 

\subsection{Kink damage results}

\begin{figure}[h]
  \centerline{\epsfxsize=8cm
                   \epsfbox{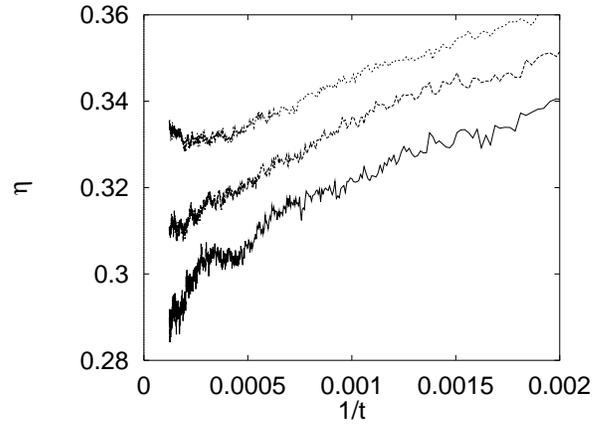}
  }
  \caption{Local slopes of the Hamming distance ($\eta$) in the A model, for 
$p=0.130, 0.132, 0.134$ (curves from bottom to top). The simulations were 
started from random initial state. 
Statistical averaging was done over $6\times 10^5$ samples.}
\label{fig1}
\end{figure}

\begin{figure}[h]
  \centerline{\epsfxsize=8cm
                   \epsfbox{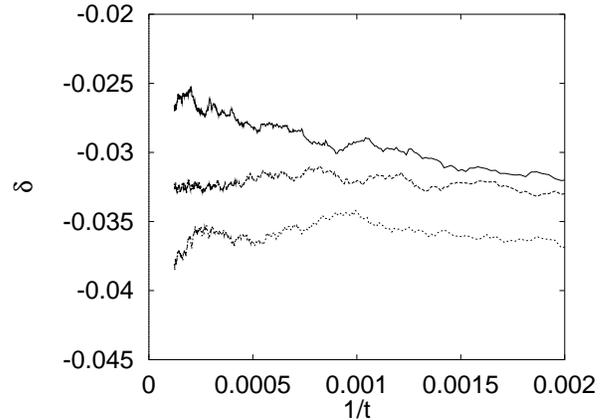}
  }
  \caption{Local slopes of the damage survival probability ($\delta$) in the A model,
 for $p=0.130, 0.132, 0.134$ (curves from bottom to top). The simulations were 
started from random initial state. 
Statistical averaging was done over $10^4$ samples.}
\label{fig2}
\end{figure}

First two replicas of lattices of the same random initial distributions but a single 
spin-flips difference condition were followed. As the local slope figure 
(Fig. \ref{fig1}) of the $D(t)$ shows the DS transition point
($p_d=0.133(1)$) seems to be slightly off the critical point ($p_c=0.1242(5)$). 
One can read off $\eta=0.31(1)$, which is close to the DP universality class value
($\eta_{DP}=0.314(3)$), but for the survival probability we got nearly zero exponent
(Fig. \ref{fig2}).

The survival probability scaling with $\delta\sim 0$ contradicts the DP scaling and
one may speculate, that we can see a finite time effect. We extended the same
time dependent simulations up to $t_{max} = 28 000$ for certain $p$ values, but there 
were no sign of change in the above results. 

\begin{figure}[h]
  \centerline{\epsfxsize=8cm
                   \epsfbox{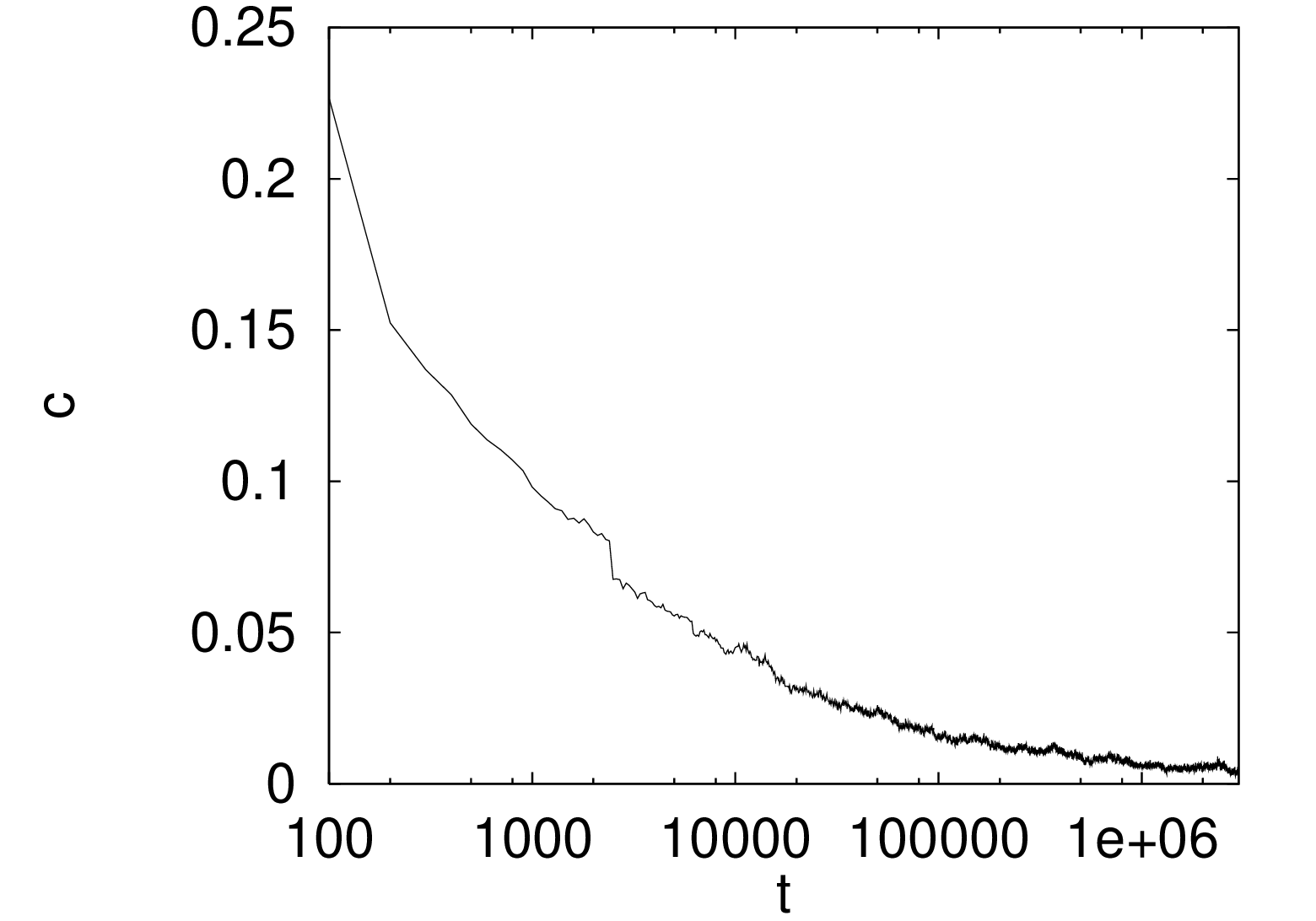}
  }
  \caption{The evolution of the kink concentration in the A model ($L=8192$) 
   started from random initial state at $p=0.124$.
  }
  \label{fig4}
\end{figure}

To check the transients the evolution of the kink concentration starting
from a disordered state have been followed on a $L=8192$ lattice. 
As the Figure \ref{fig4} shows 
there is a very long relaxation in this model, and the steady state has 
been reached following $2\times 10^6$ MC time steps only. 

\begin{figure}[h]
 \centerline{\epsfxsize=8cm
                   \epsfbox{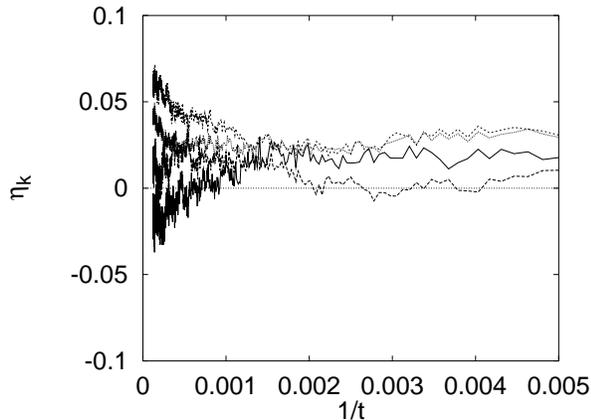}
  }
  \caption{Local slopes of the Hamming distance ($\eta_{kink}$) in the A model, 
           for $p=0.123, 0.124, 0.125, 0.126$ (curves from bottom to top) in case of
           steady state initial condition. Averaging was done over $3\times 10^5$
           independent samples.
  }
  \label{fig5}
\end{figure}

Therefore time dependent simulations from steady state initial conditions
have been performed. The initial states now were chosen to be the outcomes of runs following $5\times 10^6$ 
time steps for different $p$-s, with the usual single spin-flip difference.
Now we can see dramatical changes. First, the DS point moves to the critical
point ($p_d = 0.1242(1))$(see Fig. \ref{fig5}).

\begin{figure}[h]
  \centerline{\epsfxsize=8cm
                   \epsfbox{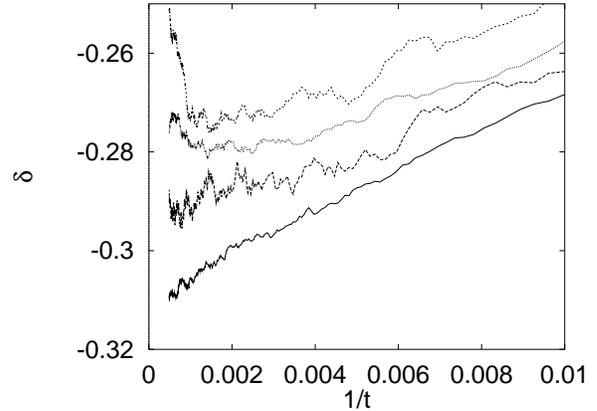}
  }
  \caption{Local slopes of the survival probability ($\delta$) in the A model, 
           for $p=0.123, 0.124, 0.1245, 0.125$ (curves from bottom to top) in case of
           steady state initial condition. Averaging was done over $50000$
           independent samples.
  }
  \label{fig6}
\end{figure}

The corresponding $\eta_{k}$ exponent is around zero, which agrees
with that of the PC universality class. 
In case of the survival probability we could use the 
conventional non-multi-spin coding algorithm with much less statistics.       
Still one can read off the same transition point with the value $\delta 
\sim 0.285(8)$ (Fig. \ref{fig6}), which is again in the PC class.

Thus we can see the emergence of PC behaviour, which is in accordance with the 
BAWe conservation of kink-damage variables and the $Z_2$ degeneration of the 
absorbing state arises from the fact that $p_c$ and $p_d$ coincided. Note, that 
the statistical errors are larger now than in the B model case, when the DS 
simulations were carried out not in the immediate neighbourhood of the critical point. 

\subsection{Spin damage results}

The parity of the spin damage variables is not conserved in this case.
When the simulations were started from random initial states we obtained the
same DS transition point as in case of the kink-damage case, but with neither DP
nor PC universality class values.  The simulations have been done both with 
conventional and multi-spin code algorithm. These resulted in the following results
for the spin damage Hamming distance as shown on figure Fig. \ref{fig9}:
\begin{figure}[h]
  \centerline{\epsfxsize=8cm
                   \epsfbox{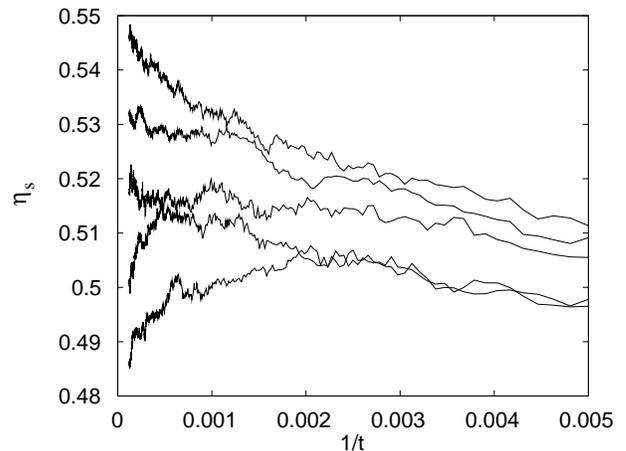}
  }
  \caption{Local slopes of the Hamming distance ($\eta$) in the A model, for 
$p=0,125, 0,127, 0.130, 0.132, 0.134$ (curves from bottom to top).
The simulations were started from random state.
Statistics over $100000 - 500000$ samples.}
\label{fig9}
\end{figure}

For the survival probability we obtained a nearly zero $\delta$ exponent, 
as in case of kink damage; see Fig. \ref{fig8}:
\begin{figure}[h]
  \centerline{\epsfxsize=8cm
                   \epsfbox{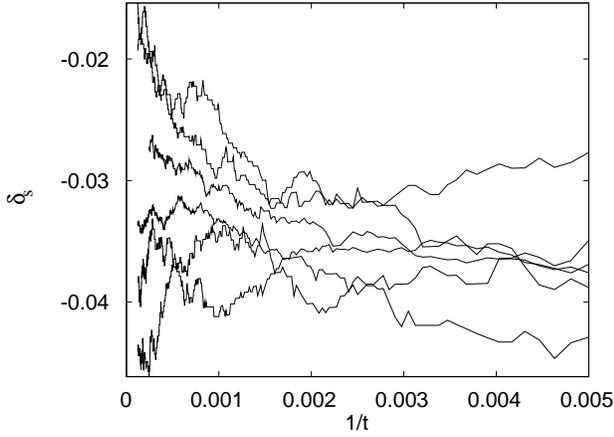}
  }
  \caption{The same as above for the survival probability.}
\label{fig8}
\end{figure}

These results are quite confusing again, especially the exponent $\eta = 0.52(1)$,
which does not belong to the 1+1 dimensional DP or to the PC class. We can not
give better explanation for this, that the very long transients prevented 
the healing of damages and the possibility to see the "true" scaling behaviour. 

\begin{figure}[h]
  \centerline{\epsfxsize=8cm
                   \epsfbox{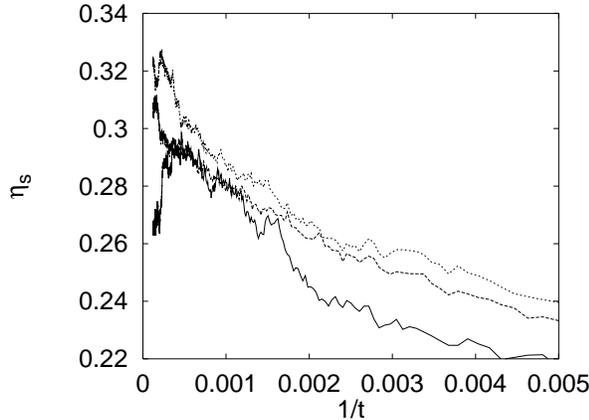}
  }
  \caption{Local slopes of the Hamming distance ($\eta_s$) in the A model, for 
$p=0,124, 0,1245, 0.126$ (curves from bottom to top).
The simulations were started from steady state. Averaging was over $3\times 10^5$ 
samples.}
\label{fig10}
\end{figure}

Indeed, if the simulations were started from near steady state
the results became very different. The DS transition point seems to coincide with the
critical point ($p_c=0.1242(5)$) and we could get a spin-damage concentration
exponent: $\eta_s=0.29(2)$ which is close to the $\eta^,=0.285(5)$ of the PC
scaling (Fig. \ref{fig10}).

\begin{figure}[h]
  \centerline{\epsfxsize=8cm
                   \epsfbox{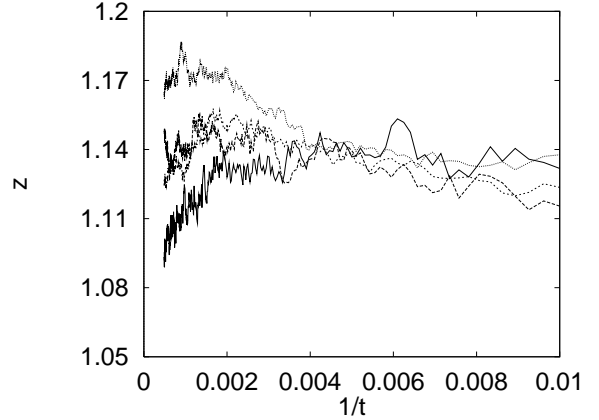}
  }
  \caption{Local slopes of $R^2(t)$ ($z$) in the A model, 
for $p=0,124, 0,1245, 0.125, 0.126$ (curves from bottom to top). 
The simulations were started from steady state. 
Averaging was done over $5\times 10^4$ independent run.}
\label{figz}
\end{figure}

The results for the survival probability ($\delta_s$) and $z_s$ coincided with
that of the kink-damage case, which can be understood by the following reasons.
Although theoretically to each spin-damage absorbing state can correspond 
two kink-damage absorbing state
(by flipping all spins of one replica), simulations showed the kink and the spin
damage died out always at the same time.
In case of the spreading one can easily check that the $R^2$ measurements should
give the same results for both kink and spin damage cases, because the beginning 
and the end of the perturbed region is almost the same. 
Indeed the simulations resulted in the same PC-like $z$ ($z=1.14(1)$)
exponent in both cases (Fig \ref{figz}).

\subsection{Finite size scaling for both cases}

\begin{figure}[h]
  \centerline{\epsfxsize=8cm
                   \epsfbox{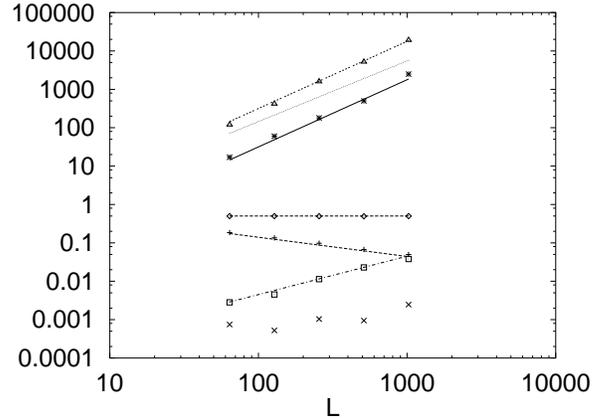}   
  }
  \caption{Finite size scaling results for the spin and kink damage for the A model.
   The diamonds correspond to the spin damage concentration, the squares to the
fluctuations of it. The crosses correspond to the kink damage concentration, the
"x"-s to the fluctuations of it. Triangles and stars denote characteristic times
$\tau$ of the spin and the kink damage cases.
Averaging was done over $500$ surviving samples.}
\label{graafss} 
\end{figure}

The finite size scaling simulations were performed at the DS transition point
($p_d = 0.1242$) for system sizes $L=64,128,...1024$. The necessary time steps 
to reach steady state were $t = 10000, 20000, ... $ respectively. 
The results can be seen on Figure \ref{graafss}. 
In case of the kink damage one can see regular PC-like scaling for the
concentration: $-\beta_k/\nu_{\perp}=-0.5$, the fluctuations of it:
$\gamma_k/\nu_{\perp}=0$ and for the critical dynamical exponent 
$Z_k = \nu_{\|} / \nu_{\perp}=1.75$.
In case of the spin damage we can see a constant $0.5$ steady state 
concentration
for all system sizes, resulting in $\beta_s/\nu_{\perp}=0$ as in case of 
the pure
Glauber Ising model at $T=0$ and the NEKIM model at the PC transition point.
In agreement with this and Fisher's static scaling law :
\begin{equation}
\gamma = d \nu_{\perp} - 2 \beta \label{FSLAW}
\end{equation}
the fluctuations of it exhibit a linear scaling law ($\gamma_s/\nu_{\perp}=1$.
Whereas the scaling of the characteristic time is described by the 
as what was found in case of the pure NEKIM model at the PC transition point
for the spin variables \cite{meod}. 

\section{The NEKIM model} 

The PC universality appears in a class of non-equilibrium dynamic Ising models,
where the kinks corresponding to '$01$' and '$10$' domain walls evolve according to
the BAWe rules \cite{Nora}. The dynamics is composed of the alternating application of
\begin{itemize}
\item a zero temperature spin flip lattice update:
\begin{equation}
w_i = {\frac{\Gamma}{2}}(1+\delta s_{i-1}s_{i+1})\left(1 -
{\gamma\over2}s_i(s_{i-1} + s_{i+1})\right)
\end{equation}
where $\gamma=\tanh{{2J}/{kT}}$ ($J$ denoting the coupling constant in
the Ising Hamiltonian), $\Gamma$ and $\delta$ are further parameters resulting in
random walk, annihilation of kinks; 
\item and a spin-exchange lattice update:
\begin{equation}
w_{ii+1}={1\over2}p_{ex}[1-s_is_{i+1}],
\end{equation}
where $p_{ex}$ is the probability of spin exchange, resulting in kink $\to$ 3 kink
creation. 
\end{itemize}
The spin-flip part has been applied using two-sub-lattice updating, 
while $L$ MC spin-exchange attempts has been done randomly 
using the outcome state of the spin-flip part. All these together
has been counted as one time-step of exchange updating. 
(Usual MC update in this last step enhances the effect of $p_{ex}$ and leads to 
$\delta_c=-.362(1)$).
In \cite{Nora,meod} we have started from a random
initial state and determined the phase boundary in the ($\delta, pex$) plane.
The phase space is composed of an active phase with free kinks and an absorbing,
ordered phase without kinks provided the initial state has an even number of kinks.
They are separated by a second order phase transition
line of PC universality. To investigate the damage spreading properties we have
chosen to fix $\Gamma=0.35, p_{ex}=0.3$ and change $\delta$  (that will play the
role of $p$ now). The PC critical point has been determined precisely 
\cite{meor}: $\delta_c=-0.395(2)$.

\subsection{Time dependent simulations}

\begin{figure}[h]
  \centerline{\epsfxsize=8cm
                   \epsfbox{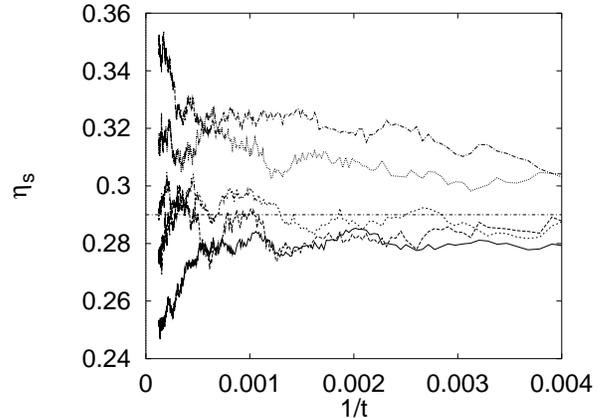}
  }
  \caption{Local slopes of the spin damage concentration ($\eta$), for 
$p=0.386, 0,39, 0.392, 0.395, 0.4$ (curves from bottom to top). 
Statistical averaging was done over 20000-40000 samples.}
\label{fig14}
\end{figure}

\begin{figure}[h]
  \centerline{\epsfxsize=8cm
                   \epsfbox{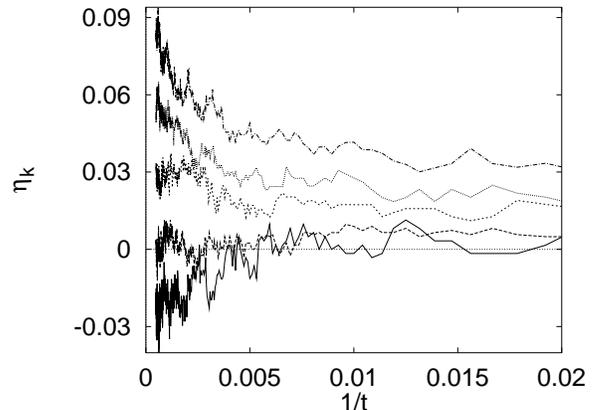}
  }
  \caption{Local slopes of the kink damage concentration ($\eta_{k}$), for 
$p=0.385, 0,39, 0.395, 0.4 0.405$ (curves from bottom to top). 
Statistical averaging was done over 100000 samples.}
\label{fig13}
\end{figure}

Time dependent simulations up to $t_{MAX}=8192$ were performed and we found that
the DS transition point coincides with $\delta_c$ within statistical errors.
The simulation results now were less sensitive whether we started from random initial
state or from steady state. For the spin damage density we obtained
$\eta_{s}=0.29(1)$ when we started from steady state, as in case of the A model. 
(If we started from random initial state $\eta_{s}=0.38(2)$ 
scaling appeared at the $\delta_c$ point). This is shown on Fig. \ref{fig14}.
The Hamming distance measurements for kinks resulted in $\eta\sim 0$ in 
accordance with the PC universality class value (Fig. \ref{fig13}).

\begin{figure}[h]
  \centerline{\epsfxsize=8cm
                   \epsfbox{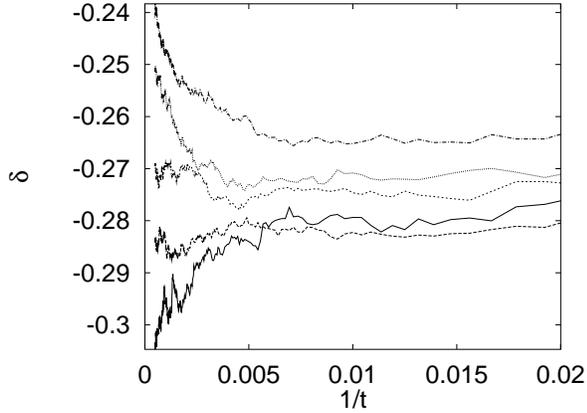}
  }
  \caption{Local slopes of the damage spin and kink survival probability ($\delta$), 
for $p=0.385, 0,39, 0.395, 0.4 0.405$ (curves from bottom to top). 
Statistical averaging was done over 100000 samples.}
\label{fig15} 
\end{figure}

\begin{figure}[h]
  \centerline{\epsfxsize=8cm
                   \epsfbox{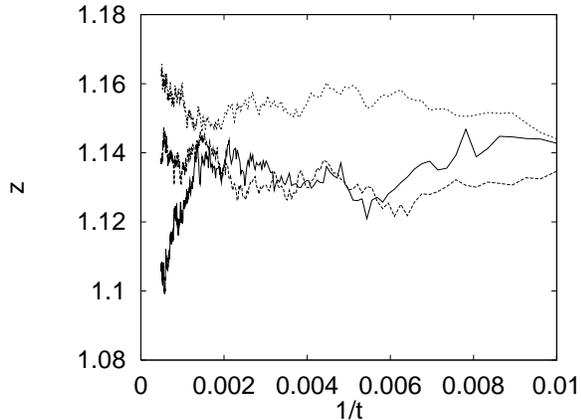}
  }
  \caption{Local slopes of $R^2$ of spin damage ($z$), for 
$p=0.39, 0.395, 0.4$ (curves from bottom to top). Statistical averaging was 
done over 100000 samples.}
\label{fig16} 
\end{figure}

The survival probability simulation gave the same results for spin and kink 
damage cases, namely they are PC-like, see (Fig. \ref{fig15}).
As in case of the A model the simulations resulted in the same PC-like $z$ 
($z=1.14(1)$) (see Fig. \ref{fig16}) exponent in both cases, 
because the spin and kink damage regions 
start and stop approximately at the same sites.

\begin{figure}[h]
  \centerline{\epsfxsize=8cm
                   \epsfbox{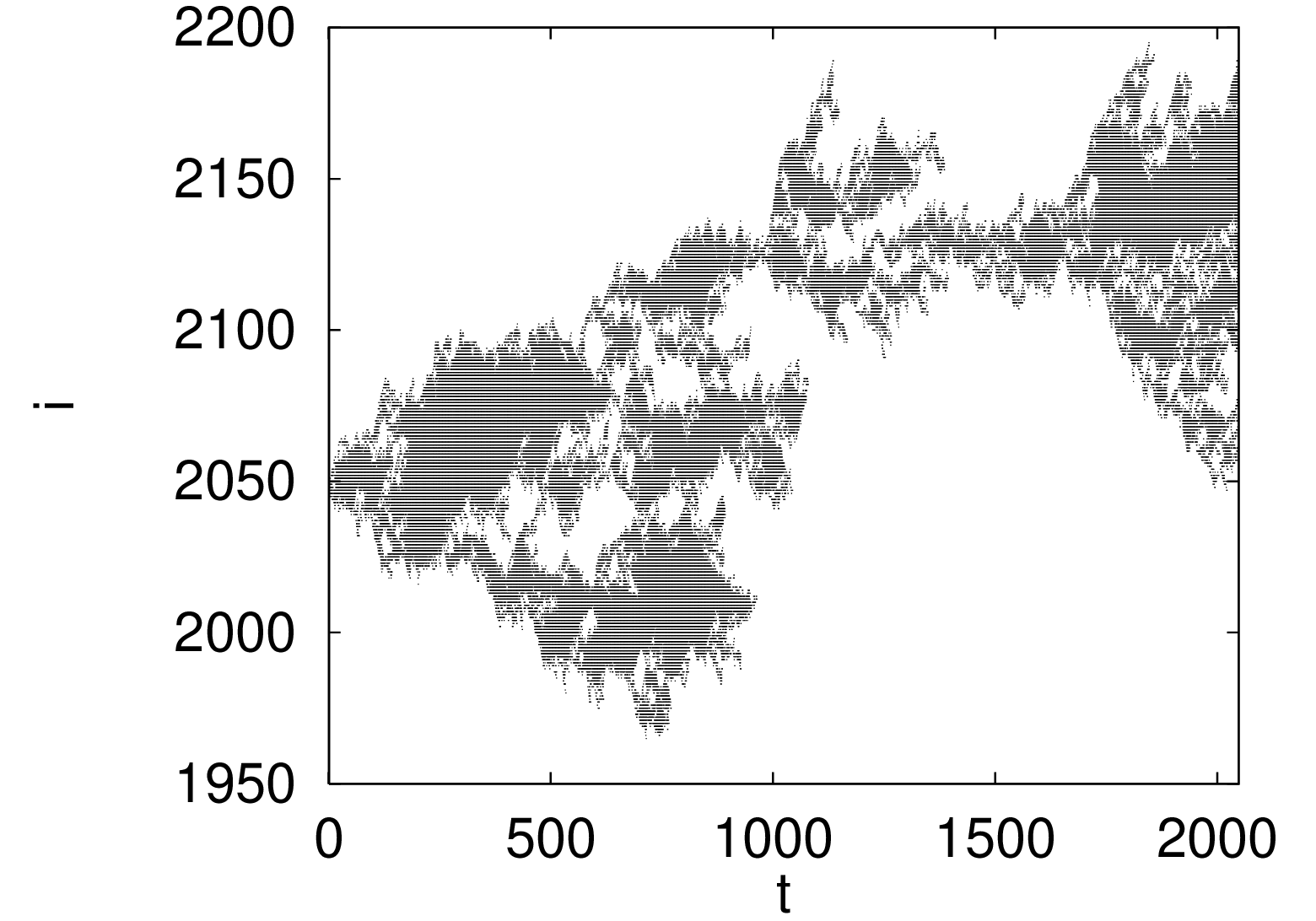}
  }
  \caption{Time evolution of the spin damage in the NEKIM model near the DS 
           transition point}
\label{damsst} 
\end{figure}

\begin{figure}[h]
  \centerline{\epsfxsize=8cm
                   \epsfbox{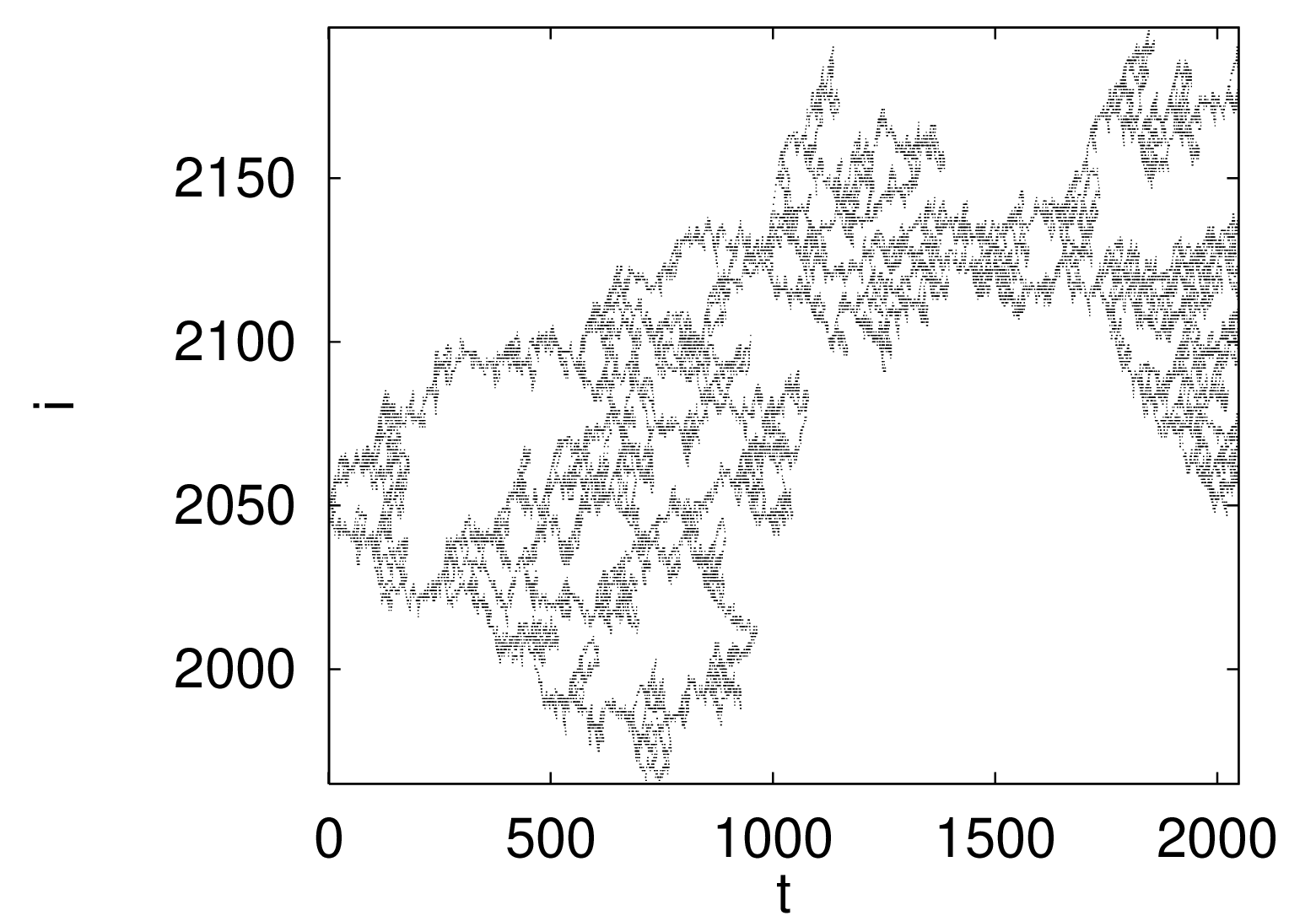}
  }
  \caption{Time evolution of the kink damage in the NEKIM model near the DS 
           transition point}
\label{damkst} 
\end{figure}

This can be observed on Figures \ref{damsst} and \ref{damkst}, where we plotted
the time evolution patterns of spin and kink damages of the same run. 
One can see that the boundaries of the perturbed regions are the same, but 
the spin-damage pattern is compact, causing the higher $\eta^,_s$ exponent.

\subsection{Finite size scaling}

\begin{figure}[h]
  \centerline{\epsfxsize=8cm
                   \epsfbox{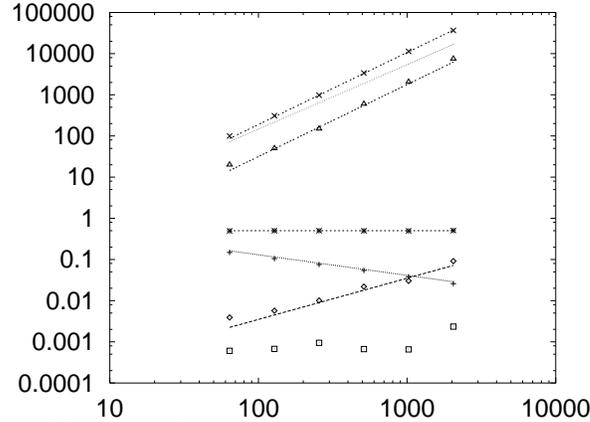}
  }
  \caption{Finite size scaling results for the NEKIM model. 
The crosses correspond to kink damage concentration, the squares to the
fluctuations of it. The stars correspond to spin damage concentration, diamonds
to the fluctuations of it. The triangles and the x-s correspond to the kink and
spin damage characteristic times ($\tau$). 
Averaging was done over $500$ surviving samples.
The intermediate dotted line shows DP like scaling law for $\tau$.}
\label{damfssNKI} 
\end{figure}

Finite size scaling simulations were performed at the DS transition point
($p_d = 0.395$) for system sizes $L=64,128,...,2048$. The necessary time steps 
to reach steady state were $t = 7000, 14000, ... ,200000$ respectively. 
Figure \ref{damfssNKI}. summarises the results.


The kink concentration shows a nice scaling with $\beta_k/\nu_{\perp}=0.5$, 
while the
fluctuations had no pronounced slope on the log-log plot, suggesting 
$\gamma_k/\nu_{\perp}=0$. Both of these values are in agreement with the PC universality
class. The spin damage concentration is constant ($c=0.50(1)$) meaning 
$\beta_s/\nu_{\perp} = 0.0$ similarly to the A-model case. A linear scaling with
$\gamma_s/\nu_{\perp}=1$ could be fitted for the fluctuation of it, meaning 
$\gamma_s = \nu_{\perp}$. This means that the exponents are in agreement with 
Fisher's scaling law (eq. \ref{FSLAW}) again.
For the critical dynamical exponent $Z = \nu_{\|} / \nu_{\perp}$ 
PC-like scaling have been obtained for both the spin and kink damage cases.
As the Figure \ref{damfssNKI} shows fitting can be done with 
$L^{1.75}$ (PC value) in both cases. The deviation from DP scaling law can 
clearly be seen on the figure too.
It is reasonable to assume that the scaling behaviours are inherited from 
the pure NEKIM model.

\section{Conclusions}

The damage spreading behaviour of three one-dimensional, non-equilibrium
models, exhibiting parity conserving phase transition has been investigated
numerically. The A SCA model of Grassberger has been found to be very sensitive
on the initial conditions of the DS simulations. Acceptable results -- which are
in good agreement with that of the NEKIM model -- can be found only when the
DS simulations are started from steady state. 

When the DS transition point coincided (within statistical error) 
with the ordinary critical point of the
model (NEKIM, A model) interesting things have happened on both the spin and 
the kink damage level.

In case of the B model the DS transition point was found to be far away from the
critical point and all exponents (on spin and kink damage level equally) show DP
universality class behaviour independently of the parity conservation of damage 
variables.

The following tables summarise the simulation results for the transition points and DS
exponents of the models investigated as well as DP and PC critical exponent
estimates from refs. \cite{PCD,meod}.

\bigskip
\centerline{Kink damage results}
\bigskip
\centerline{
\begin{tabular}{|c|c|c|c|c|c|}\hline
        & NEKIM    & GR-A      & GR-B    & DP & PC \\ \hline
$p_c$   & 0.395(5) & 0.1242(5) & 0.539(1)&    &    \\ \hline
$p_d$   & 0.395(5) & 0.1242(5) & 0.633(1)&    &    \\ \hline
$\eta$& 0.03(3)  & 0.03(3)   & 0.31(2) & 0.3137(1)& 0.0000(1) \\ \hline
$\delta$& 0.28(1)  & 0.285(8)  & 0.160(2)& 0.1596(4)& 0.285(2)  \\ \hline
$z$     & 1.14(1)  & 1.14(1)   &   -     & 1.2660(1)& 1.141(2) \\ \hline
$Z$     & 1.74(3)  & 1.75(8)   & 1.59(4)& 1.5798(2)& 1.750(5)  \\ \hline
$\beta/\nu_{\perp}$  & 0.500(6) &  0.48(2)  &  0.27(2) &0.2522(6)& 0.500(5) \\ \hline
$\gamma/\nu_{\perp}$ & 0.07(1) &  0.1(1) &  0.5 & 0.4956(2)&0.00(5)\\ \hline
univ.   & PC       &  PC       &  DP     &      &   \\ \hline    
\end{tabular}
}
\bigskip
\bigskip
\centerline{Spin damage results}
\bigskip
\centerline{
\begin{tabular}{|c|c|c|c|c|c|}\hline
        & NEKIM    & GR-A      & GR-B    & DP & PC \\ \hline
$p_c$   & 0.395(5) & 0.1242(5) & 0.539(1)&    &    \\ \hline
$p_d$   & 0.395(5) & 0.1242(5) & 0.633(1)&    &    \\ \hline
$\eta^,$& 0.29(1)  & 0.29(2)   & 0.32(2) & 0.3137(1)& 0.285(2) \\ \hline
$\delta$& 0.28(1)  & 0.285(8)  & 0.160(2)& 0.1596(4)& 0.285(2)  \\ \hline
$z$     & 1.14(1)  & 1.14(1)   &   -     & 1.2660(1)& 1.141(2) \\ \hline
$Z$     & 1.75(1) & 1.79(5)   & 1.48(9)& 1.5798(2)& 1.750(5)  \\ \hline
$\beta/\nu_{\perp}$  & 0.0001(1) & -0.001(1)  &  0.26(2)&0.2522(6)& 0.500(5) \\ \hline
$\gamma/\nu_{\perp}$ & 1.00(7) &  0.98(6)  &  0.46(4)& 0.4956(2)&0.00(5) \\ \hline
univ.   & CPC       &  CPC     &  DP     &      &   \\ \hline
\end{tabular}
}
\bigskip

The $\eta^,$ is the PC exponent when the BAWe process is started from odd 
number of particles.
The CPC notation denotes the compact version PC universality class in analogy
with the CDP in case of DP universality. 
The conclusions of this paper are in agreement with previous 
works concerning PC to DP universality changes (example \cite{ParkDP,meod,HinPC}). 
It is also in agreement with the recent DS study of Hinrichsen et. al\cite{HD},
because when they obtained PC class DS transition in case of the one-dimensional
kinetic Ising model they created a mixed dynamics, which satisfies both the
BAWe parity conservation (i.e. they followed the kinks) and the symmetric ground
state condition (i.e. they generated passive states by "switching" between two 
dynamics that results in double degeneration: no-damage versus full-damage).
In our case that has never happened that one replica would became completely
reflection symmetric to the other. The simple dynamical rules always have driven the
states to the same sector, where spin and kink damages died out simultaneously.

One may assume this scheme to be valid for non-DS dynamical transitions too,
if the model has multiple absorbing states (with or without $Z_2$ symmetry). 
This hypothesis is strengthened by simulations
of the pure NEKIM model at the PC transition point. Starting the time dependent
simulations from a single seed we measured the usual $\eta$ and $\delta$
exponents of the spins. In accordance with the DS results we obtained 
PC exponents again:  $\eta^,=0.285(5)$, $\delta=0.285(5)$. 
A detailed study of CPC for the sipns in the framework of NEKIM
and its connection to the CDP point of Domany-Kinzel CA will be published 
soon \cite{cpc}. 

These exponents satisfy the generalised hyper-scaling law \cite{PCP} with
$\beta=0$:
\begin{equation}
2 \left ( 1 + {\beta\over\beta^,} \right ) \delta^, + 2\eta^, = d z
\end{equation}
where $\beta^,$ is the ultimate survival probability exponent.
Note that the discontinuous phase transition, the compact clusters 
(see. Fig. \ref{damsst}) and the form of the hyper-scaling law suggests that
our case is the parity conserving version of the Compact Directed Percolation 
\cite{DKSCA,Essam,DickTre}.

\acknowledgements
The authors thank Haye Hinrichsen and J. F. Mendes for helpful discussions. 
Support from the Hungarian research fund OTKA (nos. 023552 and 023791) and 
from NATO grant CRG-970332 is acknowledged.
The simulations were performed partially on the FUJITSU AP-1000, and the ASTRA2
parallel supercomputers.

\end{document}